# Nanoscale mineralogy and organic structure in Orgueil (CI) and EET 92042 (CR) carbonaceous chondrites studied with AFM-IR spectroscopy

Van T.H. Phan[1], Rolando Rebois[1], Pierre Beck[1], Eric Quirico[1], Lydie Bonal[1], Takaaki Noguchi[2]

[1]Institut de Planétologie et d'Astrophysique de Grenoble (IPAG), Université Grenoble Alpes/CNRS-INSU, UMR 5274, Grenoble F-38041.

[2]Division of Earth and Planetary Sciences, Graduate School of Science, Kyoto University, Kitashirakawaoiwake-cho, Sakyo-ku, Kyoto 606-8502, Japan.

**Corresponding author:**

Van T.H. Phan

Email: thi-hai-van.phan@univ-grenoble-alpes.fr

**Abstract**

Meteorite matrices from primitive chondrites are an interplay of ingredients at the sub-µm scale, which requires analytical techniques with the nanometer spatial resolution to decipher the composition of individual components in their petrographic context. Infrared spectroscopy is an effective method that enables to probe of vibrations at the molecule-atomic scale of organic and inorganic compounds but is often limited to a few micrometers in spatial resolution. To efficiently distinguish spectral signatures of the different constituents, we apply here nano-IR spectroscopy (AFM-IR), based on the combination of infrared and atomic force microscopy, having a spatial resolution beyond the diffraction limits. Our study aims to characterize two chosen meteorite samples to investigate primitive material in terms of bulk chemistry (the CI chondrite Orgueil) and organic composition (the CR chondrite EET 92042). We confirm that this technique allows unmixing the IR signatures of organics and minerals to assess the variability of organic structure within these samples. We report an investigation of the impact of the widely used chemical HF/HCl (Hydrogen Fluoride/Hydrochloric) extraction on the nature of refractory organics (Insoluble Organic Matter, IOM) and provide insights on the mineralogy of meteorites matrices from these two samples by comparing to reference (extra)terrestrial materials. These findings are discussed with a perspective toward understanding the impact of post-accretional aqueous alteration and thermal metamorphism on the composition of chondrites. Last, we highlight that the heterogeneity of organic matter within meteoritic materials extends down to the nanoscale, and by comparison with IOMs, oxygenated chemical groups are not affected by acid extractions.

# 1. Introduction

Extraterrestrial materials are the major source of information on the processes that were at play during the formation of our planetary system. In particular, chondrites, that were accreted within a few million years after the proto-Sun formation, provide key information on processes occurring within the Solar protoplanetary disk (evaporation, condensation, transport, organo-synthesis…) as well as information on the geological evolution of the freshly accreted planetesimals (interaction with water, collisions, radiogenic heating). While chondrules and calcium-aluminum inclusions were formed at elevated temperatures, meteorite matrices are a mixture of components that span formation conditions ranging from high-temperature condensation (e.g., enstatite whiskers) to much lower-temperature (e.g., phyllosilicates). Meteorite matrices are also hosting a diversity of organic compounds ranging from small molecules containing only a few carbon atoms to macromolecular refractory organic matter (Sephton, 2002; Alexander et al., 2017; Quirico & Bonal, 2019). Meteorite matrices are thus an interplay of ingredients at the sub-µm scale, which requires analytical techniques with the sub-µm spatial resolution to unmix their composition and decipher their petrographic relations.

Infrared spectroscopy is a method that enables to probe vibrations at the molecule-atomic scale within gas, liquid or solid-state samples. Of major interest in Planetary Sciences is the fact that the strong dipole moment of – OH leads to the presence of strong IR absorptions, which have been used to identify water or hydrated minerals in planets, meteorites and asteroids. The various types of silicates can also be studied with infrared spectroscopy given the presence of $SiO_4^{4-}$ bending and stretching absorptions, which provides information on the silicate framework and permits to identify of specific mineral phase as well as amorphous silicates (Crovisier et al., 1997; Beck et al., 2014). Infrared spectroscopy is also a method that is used intensively to study the structure of carbonaceous compounds since the carbon framework can be investigated through the identification of chemical groups (e.g., C=C, C-C, C-H bonds), as well as the nature and bonding environment of heteroatoms (O, N, S) (Orthous-Daunay et al., 2013; Phan et al., 2021).

There are, however, two major challenges with the application of IR spectroscopy to natural samples. First, to be quantitative, spectra are often acquired in transmission mode, which requires an appropriate preparation for the sample to be optically thin. But given that infrared absorption strengths are variable among the species of interest, the optimum thickness may depend on the chemical compound being investigated. Second, conventional IR spectrometers are using mirror-based optics to focus the beam toward the samples, and the spatial resolution is intrinsically limited by diffraction. This means the typical spatial resolution is 5-10 µm for a sample is studied in the mid-infrared, and that in the case of chondritic matrices, the different components will be mixed within the analyzed area. This generates major difficulties in interpreting the obtained spectra since spectral congestion will occur between silicate and organically related absorptions. One possibility to circumvent this issue in the case of organic compounds is to use chemical extraction and to perform IR spectroscopy on the extracted organics. This has been done and revealed for instance chemical variabilities between IOM of different carbonaceous chondrites (Cody and Alexander, 2005; Alexander et al., 2007; Alexander et al., 2010; Orthous-Daunay et al., 2013; Quirico et al., 2014; Quirico et al., 2018). In that case, however, information on the possibility of heterogeneities of organics within a sample is lost, as well as the petrographic relation with other constituents (silicates, metals…). Also, harsh chemical extraction may have an impact on the oxygenated groups of organic matter (Painter et al., 1981; Petersen et al., 2008).

The recent development of Nanoscale IR spectroscopy offers some exciting perspectives in the field of Earth Science in general, and in the case of meteoritic studies the possibility to circumvent the aforementioned issues. This method is based on the coupling between atomic-force microscopy and infrared laser sources, which provides an opportunity to break the diffraction limit (Dazzi et al., 2005; Dazzi et al., 2012). This technique has been already applied to extraterrestrial samples (Kebukawa et al., 2018; Mathurin et al., 2019; Yesiltas et al., 2021). Here, we present results obtained using atomic force microscope – infrared (AFM-IR) spectroscopy on two raw matrix meteorites samples, and compared them with those obtained on IOM extracted by HF/HCl digestion. They were chosen to investigate primitive material in terms of bulk chemistry (the CI chondrite Orgueil) and organic composition (the CR chondrite Elephant Moraine (EET) 92042 - the weathering grade B). Our objectives with this work are:

i) assess the performance of the AFM-IR technique for natural samples by looking at reference terrestrial and extraterrestrial materials.
ii) unmix the IR signatures of organics and minerals to assess the variability of organic structure within these samples.
iii) assess the impact of the widely used chemical extraction on the nature of refractory organics.
iv) provide insights into the mineralogy of meteorites matrices from these two samples with AFM-IR.

## 2. Materials and methods
## 2.1. Sample set

We selected two meteorites including the EET 92042 CR chondrite, the Orgueil CI chondrite, as well as their extracted insoluble organic matter (IOM). The Orgueil (CI) chondrite was chosen because of its primitive bulk chemical composition (Tomeoka & Buseck, 1988; Tomeoka et al., 1989), whereas the EET 92042 sample was selected given its expected pristine nature based on its petrology (e.g., Brearley, 2006). The bulk meteorites were prepared through two different methods: focus ion beam milling and sulfur embedding ultramicrotomy, which will be described in the following part. The IOM samples of EET 92042 and Orgueil were extracted by demineralization with HF/HCl acid attacks to remove sulfates, carbonates, silicates and neo-formed fluoride ((Amari et al., 1994; Gardinier et al., 2000) following the continual flux method described in Orthous-Daunay et al. (2010).

A series of nine natural minerals were also used for the interpretation of mineralogy in carbonaceous chondrites including saponite, dolomite, kaolinite, gypsum, smectite, calcite, amorphous silicate, olivine, and antigorite (Table 1). All minerals were provided by the collection of the Observatoire des Sciences de la l'univers de Grenoble. Those are natural samples and may content some impurities (ex: Cr in olivine), but they are more suitable for their intrinsic complexity.

## 2.2. Sample preparation

The samples were prepared in different ways to be optimized for the AFM-IR and FTIR analysis following Table 1.

### 2.2.1. Focused Ion Beam milling (FIB)

A dual column focused ion beam using a Ga field emission gun and a scanning electron microscope FEG-SEM Carl ZEISS NVISION 40 system were used at Commissariat à l'Énergie atomique and Consortium des Moyens Technologiques Communs (CEA/CMTC, Grenoble, France). Firstly, platinum strips (~ 20 (long) × 2 (wide) × 5 (thick) μm) were deposited on the samples to minimize ion beam damage during the sample sectioning. Approximately 2 μm thick sections were prepared for AFM-IR analysis. During the initial extraction from a bulk sample, ion beam milling was processed at a $Ga^+$ ion beam voltage of 30 kV and beam currents ranging from 1 – 3 nA. A thin section, ~ 2 μm thickness was lifted out using an MM3A-EM micromanipulator (Kleindiek Nanotechnik) and was transferred to a Cu TEM half grid for final ion polishing. Ion currents in the range of 30 – 300 pA were used to achieve a thickness of about ~ 1 μm. A FIB section with 20 (wide) × 25 (long) × 1.5 (thick) μm was attached with the Cu grid. To perform AFM-IR analysis, the FIB - Cu grid was deposited on ZnS substrate and under the microscope, we separated the FIB section from Cu grid using a Tungsten needle (Fig. S-1, Supporting Information).

### 2.2.2. Sulfur embedding ultramicrotome

The fragment of Orgueil chondrite was prepared using ultramicrotomy after sulfur embedding at Kyushu University (Fukuoka, Japan), following a procedure optimized and developed by Noguchi et al. (2020). Ultrapure sulfur powder ($S_8$) (99.998%, Sigma – Aldrich) was put down to a stainless-steel stub located at a heated plate and heated to 170-180 ºC. We checked the quality of this sulfur droplet through its transparence, before inserting the Orgueil grain into it. All the detailed procedure was checked under the microscope before setting into the ultramicrotome as described in Noguchi et al. (2020). Before slicing sample sections, the sulfur droplet was cut in four edges using a DIATOME diamond knife trim 45º. The meteoritic fragment was sliced into ~ 90 – 100 nm ultrathin sections with a Leica ultramicrotome using a DIATOME ultra 35º diamond knife. After cutting, the released sections were deposed at the surface of deionized ultrapure water, filled on the curve of the diamond knife, and then transferred to a ZnS substrate. Before AFM-IR measurement, sections were slightly heated at 70 -80 ºC in the oven during the night to expel sulfur.

*2.2.3. Pressed and KBr pellet preparation*

For µ-FTIR analysis, the fragments of IOM/meteorites/minerals (~ 50 µm × 50 µm) were pressed between two diamond windows (type IIa diamond, 3 mm diameter, 500 µm thick) to obtain thin and flat samples (1 – 5 µm thick). The upper window was removed to avoid interference during the measurements in transmission mode. This sample preparation strongly reduces scattering artefacts and is well suited for spectral imaging. The pressed samples were not only used for µ-FTIR but also for optimization of the measurement's conditions of AFM-IR analysis (e.g., power, thickness), allowing to maximize the signal to noise ratio (SNR) of AFMIR spectra.

For the FTIR measurements of 9 reference minerals (Table 1), mineral powders (0.2 ± 0.02 mg) were ground and mixed with 300 mg KBr with a MM200 Retsch grinder equipped with an agate mortar filled with a single ball (30 Hz, 30 min). Then, 13 mm diameter pellets were prepared using a press Atlas$^{TM}$ Manuelles (Eurolabo), under 400 bars pressure and mild heating (60 ºC), then cooled down for 30 min.

For the AFM-IR analysis of pure minerals, 13 mm double-layer pellets were prepared using a first 700 mg of KBr pellet topped with 100 mg of minerals and then pressed again. The pressure was applied under pumping and gentle heating to minimize moisture in the hygroscopic KBr.

## 2.3. Conventional FTIR spectroscopy

The FTIR instrument used in this study is a Bruker Hyperion3000. The beam is modulated in an interferometer and the intensity is collected by a liquid nitrogen-cooled – mercury cadmium telluride (MCT) detector. The spectral range was 4000 – 400 cm$^{-1}$ with a spectral resolution of 4 cm$^{-1}$ and 200 scans to get satisfying SNR. All measurements were done using a cell connected to a turbomolecular pump assisted with a membrane primary pump, with an optical path through two ZnS windows (Beck et al., 2010). The loaded diamond window was maintained to a secondary vacuum (typically ~ 10$^{-6}$ mbar) and slightly heated to 80 °C to remove weakly bonded water from the sample. Background measurements were performed from an area next to the sample to monitor window cleanliness and reference baselines were acquired immediately before each sample measurement.

## 2.4. AFM-IR spectroscopy and imaging

The IR absorption spectra and maps were collected using a NanoIR3s$^{TM}$ (Brucker) at the Institut de Planétologie et d'Astrophysique (IPAG, Grenoble, France), which includes an Atomic Force Microscope (AFM) that can scan the sample and generates topographic images at nm vertical resolution. In addition, the nanoIR3s consists of two pulsed and tunable IR sources to excite infrared absorption in a sample (Bruker, 2019). As the sample heats up by absorbing radiation, the local rapid thermal expansion afterwards is detected by an AFM probe. The fast expansion leads the AFM cantilever to oscillate at its contact resonant oscillations (Dazzi & Prater, 2017). The amplitude of the generated oscillations is a proxy of the absorption of the sample at the excitation wavelength, and is similar to the standard transmission spectrum collected by conventional FTIR. As a consequence, the spatial resolution is not limited by the diffraction limit as conventional IR microscopes and spectra can be located with a spatial scale < 100 nm (Dazzi & Prater, 2017).

The IPAG NanoIR3s platform is equipped with two laser sources: a Firefly laser in the range 2000 to 3600 $cm^{-1}$ wavenumber and an APE Mid-IR laser Carmina in the range 600 – 2000 $cm^{-1}$ wavenumber. The APE laser sets new standards in resolution and tuning range due to its OPO/DFG architecture with four configurations and a power higher than 10 mW at 1600 $cm^{-1}$. The system can perform different AFM-IR analysis modes including tapping, contact mode and tapping AFM-IR mode. The tapping AFM-IR is based on principles from multifrequency AFM in which combine flexural eigenmodes of the AFM cantilever are excited to enhance the sensitivity of the probe-sample interaction (Mathurin et al., 2018). As a consequence, the highest signal-to-noise performance is obtained during the movement of cantilever oscillations (Dazzi et al., 2012; Mathurin et al., 2018). Each spectrum was obtained at selected points with a wavenumber spacing of 4 $cm^{-1}$ using both contact and tapping modes. The incident laser powers were set to 2 – 11% in the range 900 – 2000 $cm^{-1}$ wavenumber and 1 – 5% in the range 2000-3600 $cm^{-1}$. The IR spectra were optimized to a constant laser power level at each wavenumber with the laser power detected by an IR sensitive photodetector.

For the IR maps, the laser power was set to 25 - 80% in the range of 900 – 2000 $cm^{-1}$ with the tapping AFMIR mode to avoid mechanical damages from the probe at the sample surface. All AFM scans in the AFM-IR mode were done using a semi-tapping gold-coated cantilever model PR-EX-TnIR-A-10 (resonance frequency: 75 ± 15 kHz, spring constant: 1 – 7 N/m) which allows working with both on Contact and Tapping AFM-IR modes.

## 3. Results

### 3.1. AFM-IR of pure reference minerals

The fundamental vibrational modes of nine minerals including dolomite, kaolinite, gypsum, saponite, smectite, calcite, antigorite, amorphous silicate and olivine in dunite were assigned to the bands displayed in Mid-IR spectra, from both in µ-FTIR and KBr pellet FTIR (Fig. S-2). The AFM-IR spectra of calcite and antigorite were collected only using the Firefly laser in the range from 2000 – 3800 $cm^{-1}$, whereas the amorphous silicate and olivine in dunite were only measured in the range 700 – 2000 $cm^{-1}$ using the APE laser (Fig. S-2).

Dolomite and calcite (Ca-carbonate) often occur with a significant amount in CI chondrites (Tomeoka & Buseck, 1988; Tomeoka et al., 1989). AFM-IR single-point spectra in the 3-µm region reveal bands at 3006 and 2890 $cm^{-1}$ in dolomite, and 2974 and 2870 $cm^{-1}$ in calcite, corresponding to the antisymmetric and symmetric stretch vibration modes of the carbonate ion, respectively, consistently with the µ-FTIR and KBr-pellet spectra (Fig. S-2). Note that KBr-pellet spectra show the presence of $H_2O$ and an organic contaminant present in the organic matter in the KBr powder (Fig. S-2 a,c). The dolomite AFM-IR spectra also show the fundamental carbonate bands at 1806 $cm^{-1}$ ($v_1$), ~ 1450 $cm^{-1}$ ($v_3$), 880 $cm^{-1}$ ($v_2$) and 730 $cm^{-1}$ ($v_4$), with positions similar to those in conventional FTIR spectra. The AFM-IR spectra of kaolinite display features between 3600 – 3700 cm$^{-1}$ corresponding to the stretching vibration of – OH groups which are very similar to those in FTIR spectra (Fig. S-2b). The bands of 3695 and 3620 cm$^{-1}$ derived from AFM-IR spectra are lightly broader but similar to the FTIR ones corresponding to the – OH stretching modes. The features at 1110, 1040 and 920 cm$^{-1}$ are due to Si-O stretching, bending modes and – OH deformation mode in Al-OH of kaolinite structure. Overall, the AFM-IR Kaolinite spectra look very similar to FTIR spectra except for the broader and weaker feature at 1110 cm$^{-1}$.

Saponite is often interlayered with serpentine (e.g., chrysotile, lizardite and antigorite), is one of the important products in the alteration of the CI, CM and CR chondrites (Glavin et al., 2018). As seen in Fig. S-2c and h, the AFM-IR spectra of saponite and antigorite heated at 500 ºC collected in the range 3800 – 2700 cm$^{-1}$, shows the features at 3670 and 3690 cm$^{-1}$, respectively, which correspond to the OH stretching vibration. In particular, antigorite AFM-IR spectra are fully consistent with µ-FTIR spectra, whereas, for saponite, a higher intensity of the 3670 cm$^{-1}$ features is observed. In the case of the amorphous silicate and olivine in dunite, spectra were only measured in the 2000 – 700 cm$^{-1}$ range. The main bands in olivine spectra are observed at 990, 890 and 840 cm$^{-1}$ and are assigned to Si-O stretching modes. In contrast, the AFM-IR spectra of the amorphous silicate show a Si-O band with three main peaks at 1420, 1180 and 980 cm$^{-1}$, while a unique and broadband at 1000 cm$^{-1}$ due to Si-O stretching modes observed for both KBr-pellet and µ-FTIR. The reason for this discrepancy is being investigated.

Gypsum ($CaSO_4$) or the other forms of Ca-sulfate was also found in some meteorites where it was intimately associated with $CaCO_3$ such as EETA 79001 (Gooding et al., 1988). In comparison to KBr-pellet and µ-FTIR spectra in Fig. S-2d, AFM-IR spectra of gypsum reveal the presence of the stretching vibration mode of OH bond at 3600 cm$^{-1}$, and a broader peak at 3400 cm$^{-1}$ slightly different from the mid-infrared spectra centered at 3550 and 3400 cm$^{-1}$, respectively. The two deformation vibration bands of the OH groups at 1675 and 1621 cm$^{-1}$ are similar in AFM-IR and FTIR spectra. In contrast. the typical vibration bands of sulfate centered at 1110 and 1200 cm$^{-1}$ in AFM-IR spectra are different than those in FTIR spectra. One possible explanation is that upon exposure to dry air the sulfate was dedicatedly leading to some changes in the crystallographic structure.

### 3.2. Nano-scale IR spectra of Insoluble Organic Matter

In the following paragraphs, we present the AFM-IR spectra and maps obtained on insoluble organic matter extracted from the two meteorite samples.

#### 3.2.1. EET 92042 (CR) IOM

The IR absorption images of IOM extracted from EET 92042 were recorded with the IR source tuned to 1720 cm$^{-1}$, 1600 cm$^{-1}$, 1450 cm$^{-1}$ and 1040 cm$^{-1}$ corresponding to carbonyl (C=O), aromatic (C=C), aliphatic ($CH_2$ bending) and silicate groups (Si-O stretching), respectively (Fig. 1 a-e). For each map, the height topography was recorded to monitor for the thermal drift during the experiment. The infrared maps are shown in Figure 1 and reveal variability in the AFM-IR intensity over the analyzed area. The comparison to the topographic map reveals that some of the trends are related to sample height: for instance, the area with fractures or the topographic low height around points S2-S3 show overall a lower AFM-IR intensity. The contact resonance frequency is likely different in these areas, which leads to a lower AFM-IR signal (the pulse repetition rate is optimized only once before a chemical map is obtained), as well as to the fact that variable thickness can lead to variation in magnitude of the laser-induced thermal expansion. Overall, this map reveals the presence of correlated organic matter absorption at 1720 cm$^{-1}$, 1600 cm$^{-1}$ and 1450 cm$^{-1}$ and negligible signatures of silicate minerals that would produce absorptions around 1000 cm$^{-1}$ (Si-O stretching modes). These maps suggest that the IOM is quite homogenous and that minerals were efficiently removed by the chemical extraction. It also emphasizes the general need to work with smooth and flat surfaces for AFM-IR measurements (with a typical standard deviation of the topography below 50 nm).

The functional C groups present in the IOM of EET 92042 were reported in previous studies based on µ-FTIR (Kebukawa et al, 2011; Orthous-Daunay et al., 2013). The µ-FTIR spectrum of Fig. 1g (in blue) is similar to the IOM AFM-IR spectra with the absorption band peaking at 1700 and 1600 cm$^{-1}$ attributed to C=O and C=C and the $I_{1700}/I_{1600}$ ratio of 0.9. A line of AFM-IR single-point spectra (marked in Fig. 1a-b) was collected. Spectra of S1, S2 and S3 represent a spectral blank that corresponds to the ZnS substrate. The shape of the other AFM-IR spectra (e.g. S4 – S25) is quite similar to typical IOM. To compare with µ-FTIR spectra, the average spectrum, shown in Fig. 1g, based on around 100 single-point spectra is suggesting that the $I_{1720}/I_{1600}$ is around 0.9 to 1.1 which is slightly higher than the one derived from FTIR proxy (0.9).

### 3.2.2. Orgueil (CI) IOM

The AFM-IR maps of the IOM extracted from Orgueil are shown in Fig. 2. Infrared maps were recorded at the same laser wavenumbers as for the IOM of EET 92042. Again, we see a strong correlation between IR maps obtained at 1720 cm$^{-1}$, 1600 cm$^{-1}$ and 1450 cm$^{-1}$ (Fig. 2b-c). In contrast to the IOM of EET 92042, the IOM of Orgueil is not homogeneous, with an area showing a significant absorption at 1000 cm$^{-1}$ that may result from the uncompleted mineral removal through the chemical extraction or from the presence of residual SiF$_4$ crystals. This absorption is not observed on µ-FTIR spectra but this difference may be attributed to heterogeneity in the distribution of these mineral residues within the IOM (Fig. 2f).

The point AFM-IR spectra, obtained outside of the area with the strong 1000 cm$^{-1}$ signal in IR maps, reveal the presence of bands at 1700 and 1600 cm$^{-1}$ assigned to C=O and C=C, respectively, as well as at 1450 cm$^{-1}$ attributed to CH$_2$ (Fig 2g). Spectra obtained in the area with the strong 1000 cm$^{-1}$ signal in IR maps display additional signatures around 1250 and 1000 cm$^{-1}$ that are tentatively assigned to a mineral phase and/or SiF$_4$ crystal which may remain and/or form from the chemical extraction using hydrofluoric acid (HF).

### 3.3. Nano-scale IR spectra of raw carbonaceous chondrites

#### 3.3.1. EET 92042 CR chondrite

An area of a FIB section of EET 92042 was mapped with the Tapping AFM-IR mode (5 × 5 µm$^2$). Results are shown in Fig. 3. The AFM-IR images at different wavelengths were obtained at 1720 cm$^{-1}$, 1600 cm$^{-1}$, 1450 cm$^{-1}$ and 1000 cm$^{-1}$ corresponding to C=O, polyaromatic C=C, carbonate (CO$_3$) and Si-O stretching, respectively (Fig. 3b-e). These maps and the generated composite RGB images enabled the identification of three different types of domains in the mapped area (Fig. 3f). The first region is located at the bottom left of the images and can be distinguished by a strong absorption at 1450 cm$^{-1}$ and attributed to the presence of carbonate. The second region is in the form of a vein (red in the RGB image Fig. 3f) that has stronger absorption related to organics (1720 and 1600 cm$^{-1}$). The last region corresponds to the rest of the image, where the most intense signal detected can be attributed to silicates (1000 cm$^{-1}$).

Several AFM-IR spectra were obtained across this FIB section (Fig. 3g). These spectra reveal the strength of AFM-IR spectroscopy to unmix spectral signatures at a sub-µm scale since such an area (5 × 5 µm$^2$) would correspond to only one pixel using conventional micro-FTIR. The spectra obtained in the center of the carbonate grain (P19) shows signatures at 1460 cm$^{-1}$ and 870 cm$^{-1}$ together with a faint feature around 1800 cm$^{-1}$ all suggestive of carbonate. Spectra obtained in the green area of the RGB map reveal the presence of a broad and variable feature around 1000 cm$^{-1}$ that is attributed to phyllosilicates whose nature will be discussed in section 4.2. In the case of the red vein in the RGB maps, organic signatures at 1720 and 1600 cm$^{-1}$ are visible in the point spectra. This organic-rich area is finely mixed with phyllosilicate given the presence of associated silicate bands.

#### 3.3.2. Orgueil CI chondrite

IR absorption maps of a microtome section of Orgueil were obtained over an area of 4 × 4 µm$^2$, at 1720 cm$^{-1}$, 1450 cm$^{-1}$, 1150 cm$^{-1}$ and 1040 cm$^{-1}$ respectively (Fig. 4a-e). The AFM-IR RGB composite map shown in Fig. 4f is an overlay of the three channels of 1710, 1450 and 1000 cm$^{-1}$.

These maps reveal the nanoscale heterogeneity of the mineralogy and organic distribution in Orgueil, for which three spectral endmembers were identified based on point measurement. The first endmember is phyllosilicate, with a peak around 1000 cm$^{-1}$ (point O14 for example, Fig. 4), which is thinner than the silicate absorption in EET 92042 (CR). The second endmember shows a triplet at 1150, 1090 and 980 cm$^{-1}$ as well as a doublet at 1465 and 1425 cm$^{-1}$ (point O01 for example). This endmember is interpreted by the presence of a hydrated sulphate. The last endmember is attributed to organic compounds. It seems to be present diffusively in the sample except for a few areas like point O19 which shows a stronger signature around 1700 cm$^{-1}$ than the rest of the sample (Fig. 4). More spectra of organic matter were observed in the section nearby the one in Fig. 4 (e.g., S10, S13, S14 and S15 in Fig. S-3) which show clearly the signature of carbonyl and polyaromatic around 1700 and 1600 cm$^{-1}$, respectively.

## 4. Discussion

### 4.1. Organic signatures and structure, and the effect of acid extraction

Fig. 1g displays a conventional μ-FTIR spectrum of the IOM sample of EET 92042 and an AFM-IR average spectrum. The bands at 1720 (C=O) and 1600 cm$^{-1}$ (aromatic) are observed in both spectra, but the intensity of the C=O band is systematically higher than that of the aromatic band in the AFM-IR spectrum. In the AFM-IR average spectrum, the baseline (zero-absorption continuum) is not located on the side of high wavenumber: this is particularly visible at the edge of the 1720 cm$^{-1}$ band and the range 1500-900 cm$^{-1}$ is noisy. The AFM-IR average spectrum does not display the CH$_2$ and CH$_3$ bending modes (1450 cm$^{-1}$ and 1380 cm$^{-1}$) and, nor the broad congested bands at ~1250 cm$^{-1}$ (C-O, C-C, OH, etc.) that systematically appears in conventional μ-FTIR spectra, from this study and in published data (Kebukawa et al., 2011; Orthous-Daunay et al., 2013). Similar observations are achieved in spectra of Orgueil IOM regarding the ratio of the 1720 and 1600 cm$^{-1}$ bands (Fig. 2). However, AFM-IR and μ-FTIR spectra look more similar in the range 1500-900 cm$^{-1}$. The CH$_2$ and CH$_3$ bending bands are present, and the broad congested band looks more similar as well, except for two bumps at 1240 and 1000 cm$^{-1}$. These bumps could be assigned to aromatic-ethers aromatic-O-CH$_2$ and C-O bonds, respectively, and their spatial variations may point to chemical heterogeneity of Orgueil-IOM.

These spectral differences between conventional μ-FTIR and AFM-IR spectra are not due to instrumental artefacts, as we collected AFM-IR spectra of simple polymer films (e.g. polybutadiene) similar to μ-FTIR spectra (Fig. S-4). They point to the extreme sensitivity of the technique to the nature of the sample and its preparation. Natural samples are heterogeneous and porous at the nm scale, resulting in complex heat transfer. Therefore, the volumetric expansion triggered by the laser could not be controlled by the sole absorption properties of the organic material. The sample preparation also plays a key role, and spectra can depend on thickness and compaction (if it is prepared by crushing). In this respect, the interpretation of AFM-IR spectra of natural materials needs to be based on AFM-IR spectra obtained on reference samples, if possible prepared with similar sample preparation.

AFM-IR measurements on raw matrix samples of EET 92042 display signatures of minerals and organics. The 1700-1600 cm$^{-1}$ spectral region is possibly contaminated by the bending mode of molecular water. Selecting spectra with a low Si-O band minimizes this spectral contamination and reveal the actual signatures of organic matter (Fig. 5). We observe two bands at 1720 and 1600 cm$^{-1}$, with spatial variability of their relative intensities. In the CR chondrite Renazzo, organic matter is mostly present as grains of a few hundreds of nm that are not intermixed with hydrated minerals (Le Guillou et al., 2014). In the very primitive CR MET 00426, organic matter is present as individual particles over a broader range of a size (from a few nm to several µm), but the main population is around 100-500 nm in size, which is larger than the spatial resolution achieved with AFM-IR (Le Guillou et al., 2014). A diffuse organic matter is also reported, showing a higher abundance of – COOH groups and pointing to the presence of Soluble Organic Matter (Sephton, 2002). These observations have been extended to other weakly altered CR chondrites, including EET 92042 (Changela et al., 2018). Our results show no major enhancement of the C=O band, and in this respect, IOM appears as the main organic form in the probed area with little or negligible contribution from soluble organic matter. Our spectra arise from the excitation of rather large IOM grains, nevertheless, we could measure pure organic area. Last, the presence of fairly separated features at 1720 and 1600 cm$^{-1}$ in spectra obtained on raw matrix, similar to those in IOM spectra, shows that the HF/HCl acid extraction protocol has little effect on the carbonyl oxygenated groups. These groups are known to be very sensitive to acids, and in demineralized coals, they are harshly modified (Painter et al., 1981; Petersen et al., 2008). IOM samples extracted from chondrites by acid digestion are then well chemically representative of the actual refractory organic matter in the bulk chondrite.

In the case of Orgueil, spectra displaying organic features are scarcer than in EET 92042. This may be because organic matter is present in a more diffuse form, with tight intermixing with phyllosilicates (Pearson et al., 2002; Le Guillou et al., 2014a). The few spectra displaying a low silicate band reveal, however, two distinct bands at 1720 and 1600 cm$^{-1}$. The C=O band is more intense and shifted in comparison to Orgueil IOM, suggesting a contribution of SOM in the spectra obtained on the matrix. Note, however, that only a small number of AFM-IR spectra displaying exploitable organic features derived from organic matter could be collected.

**4.2. Silicate mineralogy**

*4.2.1. Phyllosilicates in Orgueil*

Unlike most carbonaceous chondrites for which chondrules or chondrule relicts are abundant, the mineralogy of Orgueil and CI chondrites is dominated by a fine-grained, dark, matrix (Tomeoka & Buseck, 1988). The mineralogy of the Orgueil matrix has been investigated using a variety of techniques (XRD, Mossbauer, TGA…), as well as in situ techniques using a combination of SEM and TEM. In the seminal work of Tomeoka & Buseck (1988), it was shown that the matrix of Orgueil is dominated by phyllosilicates, in the form of small grains (<1 µm) and larger aggregates. The mineralogy of these phyllosilicates was found to be dominated by mixtures of serpentine and smectite (saponite) as evidenced by the alteration of domains with 7 Å and 10 Å spacing. This mineralogy is consistent with bulk X-ray diffraction patterns (King et al., 2015). Infrared spectra of Orgueil obtained on bulk samples using FTIR have been reported previously (Sandford, 1986; Beck et al., 2014). These measurements revealed the presence of an absorption band centered around 10 µm together with smaller features around 1100 and 1150 cm$^{-1}$, that were previously attributed to sulfates (Sandford et al., 1986; Beck et al., 2014).

Here, for the first time, we can unmix the spectral signature of the phyllosilicate from that of other matrix constituents (Fig. 4f and 6a). Using AFM-IR spectroscopy, we are therefore able to obtain the "pure" signature of Orgueil phyllosilicate. This reveals a single and relatively narrow peak when compared to CM chondrites (REF) with a maximum of around 9.9 μm. Also apparent in the spectrum are signatures around 1710 and 1460 cm$^{-1}$ (e.g, AFM-IR spectra O19) that can be attributed to organic matter finely mixed with the phyllosilicate (Fig. 4f). It is not possible to tell if the organic matter is present within the interlayer space of the phyllosilicate.

The shape and position of the phyllosilicate feature in Orgueil (Fig. 6a) are somehow similar to saponite, but the spectra obtained at a longer wavelength with FTIR in our previous work are not identical to saponite (Beck et al., 2014). We also note that the water bending mode at 1630 cm$^{-1}$ is not identified in the spectra, while it can be seen in the saponite standard measured with AFM-IR. This suggests that the interlayer spaces of the saponite-like domains in our Orgueil sample are not populated by water molecules. A possibility is that the layers have collapsed (measurements are obtained under dry air) or that the interlayer space is populated with other species (i.e. organics; Pearson et al., 2002).

### 4.2.2. Hydrated amorphous silicates in the CR chondrite EET 92042

CR chondrites are usually described as amongst the most primitive carbonaceous chondrites. This is justified by particularly heavy nitrogen isotopic compositions, the high abundance of presolar grains and the overall low level of aqueous alteration compared to CM chondrites (Alexander et al., 2007; Bonal et al., 2013). Also, the mineralogy of the matrices of some CR chondrites, as studied by TEM, reveal the presence of abundant amorphous material that would not survive a significant amount of thermal processing (Changela et al., 2018; Le Guillou et al., 2015; Abreu and Brearley, 2010). This amorphous material is intermixed with sulfides, organic material and tochilinite (Changela et al., 2018).

In figures, 6b and 8, some spectral endmembers derived from AFM-IR measurements on the CR EET 92042 are shown. Some silicate-dominated regions are present, and different silicate signatures were found (blue spectra in figure 6b, green spectra in figure 8). The dominant silicate signature is a broad feature centered around 10 μm, with a width significantly broader than that of Orgueil phyllosilicate (see the Orgueil only Si-O spectrum in figure 6a). The position and overall aspect of the peak are consistent with FTIR spectra obtained on matrix grains of the same samples (Bonal et al., 2013). Again, comparison to FTIR data reveals the strength of AFM-IR to unmix the spectral signature in carbonaceous chondrites (Figure 8a). The overall shape of these spectra is consistent with an amorphous silicate phase. Interestingly, this spectrum reveals the presence of a feature around 1630 cm$^{-1}$ (Fig. 8b), which is the first direct evidence for the presence of water molecules within amorphous silicate from a CR chondrite (Beck et al., 2010)

It has been proposed that the hydration observed in amorphous silicates from CR chondrites could be related to the reaction between amorphous silicates and water (Le Guillou et al., 2015). The obtained AFM-IR spectra of EET 92042 amorphous silicate is remarkably similar to experimental products obtained by (Potapov et al., 2020) that studied the infrared properties of ice-silicate mixtures (Fig. 8b), that was progressively heated up to 200 K. As argued by Potapov et al. (2020) the interaction with water-ice grains can trap water molecules within or on the surface of amorphous silicates; the incorporated water molecules are tightly bonded to the amorphous silicates, which can help in delivering water in the inner part of the protoplanetary disks. Our AFM-IR observation on the CR chondrite EET 92042 provides evidence that such hydrated amorphous silicates were indeed present in the early Solar System. Because these water molecules are more tightly bonded than within the water-ice structure, these hydrated amorphous silicates therefore could have played a role in delivering water to the inner Solar System and then terrestrial planets.

### 4.3. Precipitates: carbonates and sulfates

#### *4.3.1. Carbonate mineralogy*

Both in EET 92042 and Orgueil, an absorption around 1400-1500 $cm^{-1}$ was observed, which we interpret by the presence of carbonates (Fig. 6). Looking at these peaks in more detail, the exact position differs in Orgueil and EET 92042, with a peak maximum around 1440 $cm^{-1}$ for the CI and around 1470 $cm^{-1}$ for the CR chondrites. This difference points toward different mineralogy of the carbonates present in both samples (Lane & Christensen, 1997). Also, while pure carbonate spectra were obtained for the CR chondrite, the carbonates signatures in Orgueil were always found in association with the sulfate signature (see the section below) (Fig. 6a and 7b).

The carbonate-rich region in EET 92042 presents a strong absorption peak at 1470 $cm^{-1}$, together with smaller features around 1780 $cm^{-1}$ and 860 $cm^{-1}$ (Fig. 6b and 8a). They do not show features at 1630 $cm^{-1}$ which rules out the presence of hydro-carbonates. The position of the peak found for carbonates in EET 92042 using AFM-IR are in good agreement with the carbonate-related feature in the spectra of a carbonate-rich matrix fragment measured with FTIR (EET92042#P19, green spectrum of Figure 8a). Comparison with carbonates spectra from the literature reveals that dolomite, calcite or siderite do not match the position of the infrared peaks obtained for EET 92042 carbonates. On the contrary, comparison to infrared spectra of aragonite provides a fair match to AFM-IR measurements of carbonates in EET 92042 (Fig. 8a).

Previous works have shown that carbonates in CR chondrites are mostly calcium-rich (Weisberg et al., 1993; Brearley, 2006; Jilly-Rehak et al., 2018) but little is known on their exact crystal structure. In the case of seawater, the nature of the $CaCO_3$ polymorph that can precipitate depends on both the temperature and Mg/Ca ratio of the solution (Morse et al., 1997). An elevated Mg/Ca ratio will favor the formation of aragonite rather than calcite (for example Mg/Ca >5 at T=8°C), while, on the other hand, for a given Mg/Ca ratio low temperature will favor the precipitation of calcite (at Mg/Ca=1, calcite will precipitate if T<20°C) (Morse et al., 1997).

Aragonite has been identified in CM chondrites, and its abundance depends on the petrological subtype (Lee & Ellen, 2008; Lee & Lindgren, 2015; Lee et al., 2014). It was found by Lee et al. (2014) that the abundance of aragonite decreases with an increasing extent of aqueous alteration, which suggests that aragonite was possibly the first carbonate to crystallize in the aqueous alteration sequence defined by Lee et al. (2014). Our possible detection of aragonite in EET 92042 may be explained by a relatively low level of aqueous alteration and precipitation from a fluid that experienced little chemical fractionation.

### 4.3.2. Which sulfate is in Orgueil?

As evidenced from the map obtained at 1150 cm$^{-1}$, the second most frequent IR signature observed in Orgueil (after phyllosilicates) is that interpreted as sulfate (Fig. 4d). Such a signature reveals absorption at 1150, 1090 and 980 cm$^{-1}$ (e.g., Fig. 7a). A broad feature around 1660 cm$^{-1}$ is also associated with these sulfate signatures (interpreted by structural H$_2$O) as well as an absorption around 1450 cm$^{-1}$ interpreted as a carbonate signature. To identify the nature of the sulfate present in Orgueil, the spectra obtained using AFM-IR were compared to literature spectra of sulfate of various compositions. While no perfect match was found, the closest spectra identified in the literature is copiapite, an iron-bearing hydrated sulfate (Fig. 7b). While Ca- and Mg-rich sulfate has been described in Orgueil in the form of large veins, the sulfate found at the very small scale in our sample of Orgueil seem to fairly match with basanite (CaSO$_4$.0.5H$_2$O) spectrum at 1145 and 1090 cm$^{-1}$ as previously described in Liu et al. (2018). It is possible that sulfate in Orgueil is the composition of epsomite and basanite which were observed fragment by SEM-EDS (Brearley & Jones, 1996). Copiapite can be found on Earth as an alteration of iron sulfide. Therefore, if indeed copiapite, the sulfate found in the fine-grained matrix of Orgueil could be a decomposition product of primary iron sulfide, that was formed during aqueous alteration on the CI parent body, or during terrestrial residence similarly to sulfate veins (Gounelle & Zolensky, 2001).

It was found by Tomeoka & Buseck (1988) that the phyllosilicate in Orgueil is intimately mixed with a Fe-, S- and Ni-rich phase. Since ferrihydrite was identified based on the selected area electron diffraction pattern, the Fe-S-Ni associate was interpreted by the presence of sulfate and Ni ion adsorbed onto the ferrihydrite. It is possible that the sulfate signature observed in the present work corresponds to adsorbed sulfate ions on a ferrihydrite surface, but we do not know yet the spectral signature of such an association.

### 5. Conclusion

We applied atomic force microscopy coupled infrared spectroscopy (AFM-IR) with the latest platform NanoIR3s to access meteoritic composition with nanometer-scale resolution within high and low altered chondrites and their IOMs. We achieved ~ 20 nm spatial resolution on taping imaging and ~ 150 nm lateral resolution on contact spectroscopy which is beyond the diffraction limit. AFM-IR analysis successfully separates the signatures of organic-mineral association: (i) unmix the spectral signature of the phyllosilicate from that of other matrix constituents and organic matter for the first time in Orgueil and observe the changes of structure in macro-molecular hydrocarbon under the chemical extraction; (ii) give more evidence of hydrated amorphous silicates present in the early Solar System through the AFM-IR spectra on the CR chondrite EET 92042; (iii) find out the exact carbonate crystal structure both in CR and CI chondrites; (iv) observe the sulfate signature in Orgueil (CI) similar to the mixture of Basanite and Epsomite. In addition, using the AFM-IR technique, we are not only able to image the functional groups of organic similar to the C K-edge XANES/STXM technique, but also access to the mineralogy similar to TEM analysis. Finally, the present work clearly shows that AFM-IR is useful technique and suitable for Ryugu asteroid samples, as well as IDPs and other primitive solar system materials.

**ACKNOWLEDGEMENTS**

The authors are grateful for the financial support of the H2020 European Research Council (ERC) (SOLARYS ERC-CoG2017_771691) and the Programme National de Planétologie (PNP) as well as the Centre national d'études spatiales (CNES) within the framework of the Hayabusa2 and MMX missions. We acknowledge Frédéric Charlot for the assistance during FIB preparation at CEA/CMTC (Grenoble, France). We also acknowledge Minako Takase, who helped to prepare the sulfur embedded microtome section for AFM-IR analysis at the microscopy platform of Kyushu University (Fukuoka, Japan).

**Supplementary data**

Supplementary data related to this article can be found in the Supporting Information part.

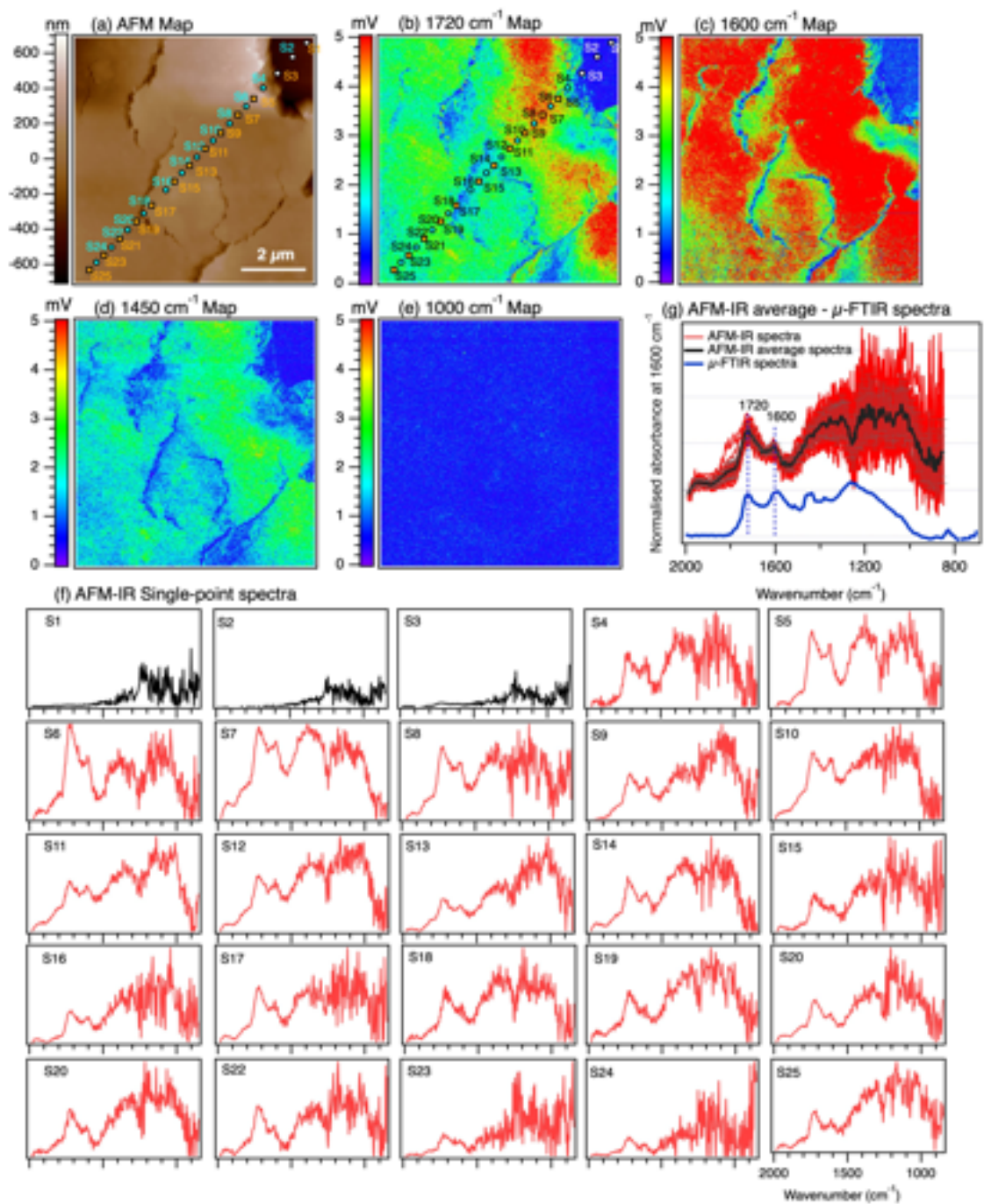

Figure 1. IOM extracted from EET 92042 (CR2 chondrites) (10 × 10 μm$^2$). IR map for each wavenumber with corresponding (a) AFM image at (b) 1720 cm$^{-1}$, (c) 1600 cm$^{-1}$, (d) 1450 cm$^{-1}$ and (e) 1000 cm$^{-1}$; (f) AFM-IR single point spectra (scaled) for the region of 2000 – 700 cm$^{-1}$ (labelled S1 to S25 with a line of 25 spectra) performed in selected locations shown on the AFM and 1720 cm$^{-1}$ image; (g) Comparison between the AFM-IR average spectra derived from 100 spectra (with the standard deviation grey bar) and μ-FTIR spectra.

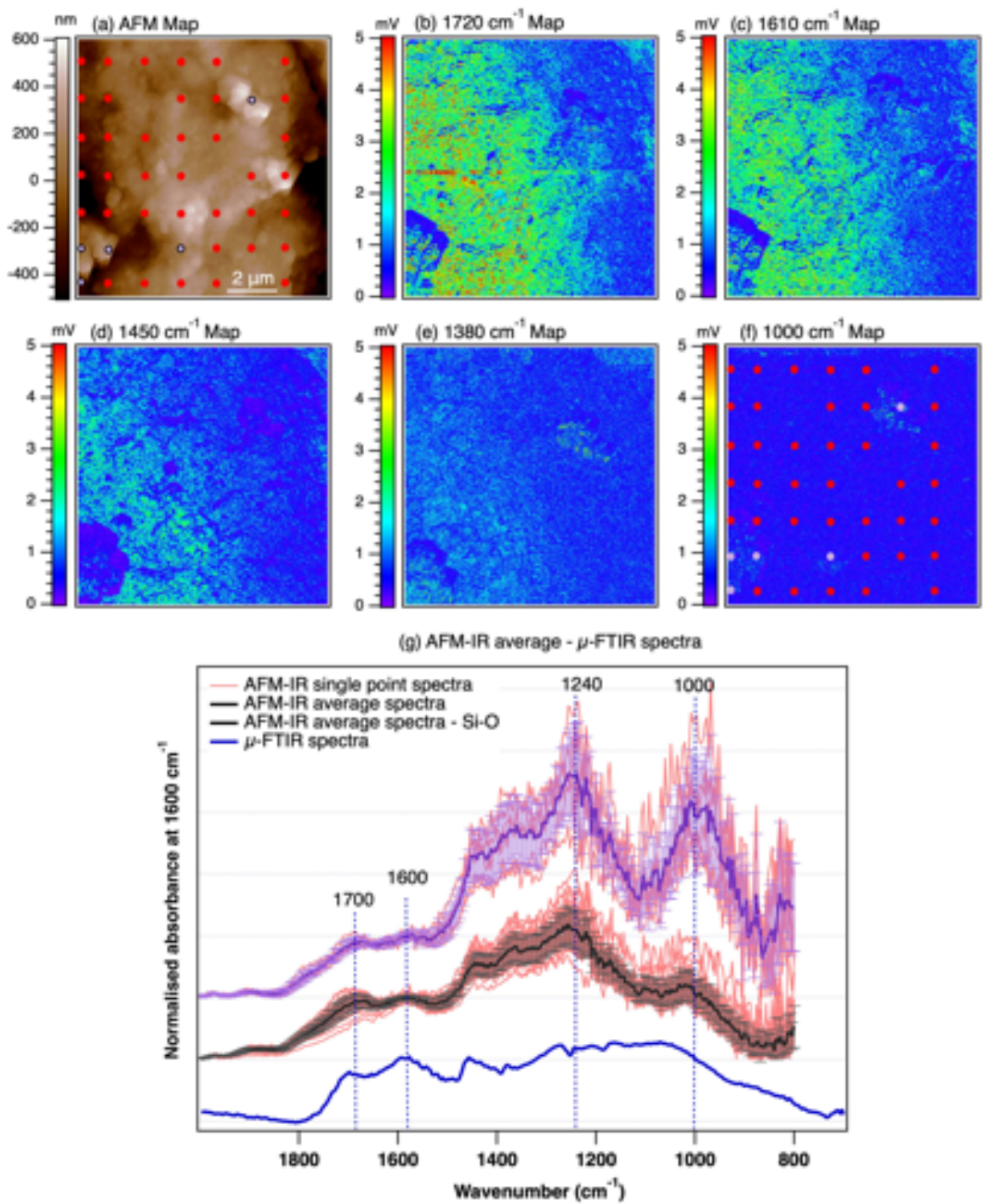

Figure 2. IOM extracted from Orgueil (CI chondrite) (10 × 10 µm²). IR map for each wavenumber with corresponding (a) AFM image at (b) 1720 cm⁻¹, (c) 1600 cm⁻¹, (d) 1450 cm⁻¹, (e) 1380 cm⁻¹ and (f) 1000 cm⁻¹. (g) Comparison of AFM-IR average spectra and µ-FTIR spectra including the 40 average (red) and 5 average (purple) spectra, including to the standard deviation bar in the position without and with the signature of 1000 cm⁻¹, respectively in the range of 2000 – 700 cm⁻¹ performed in selected locations shown on the AFM and 1000 cm⁻¹ image.

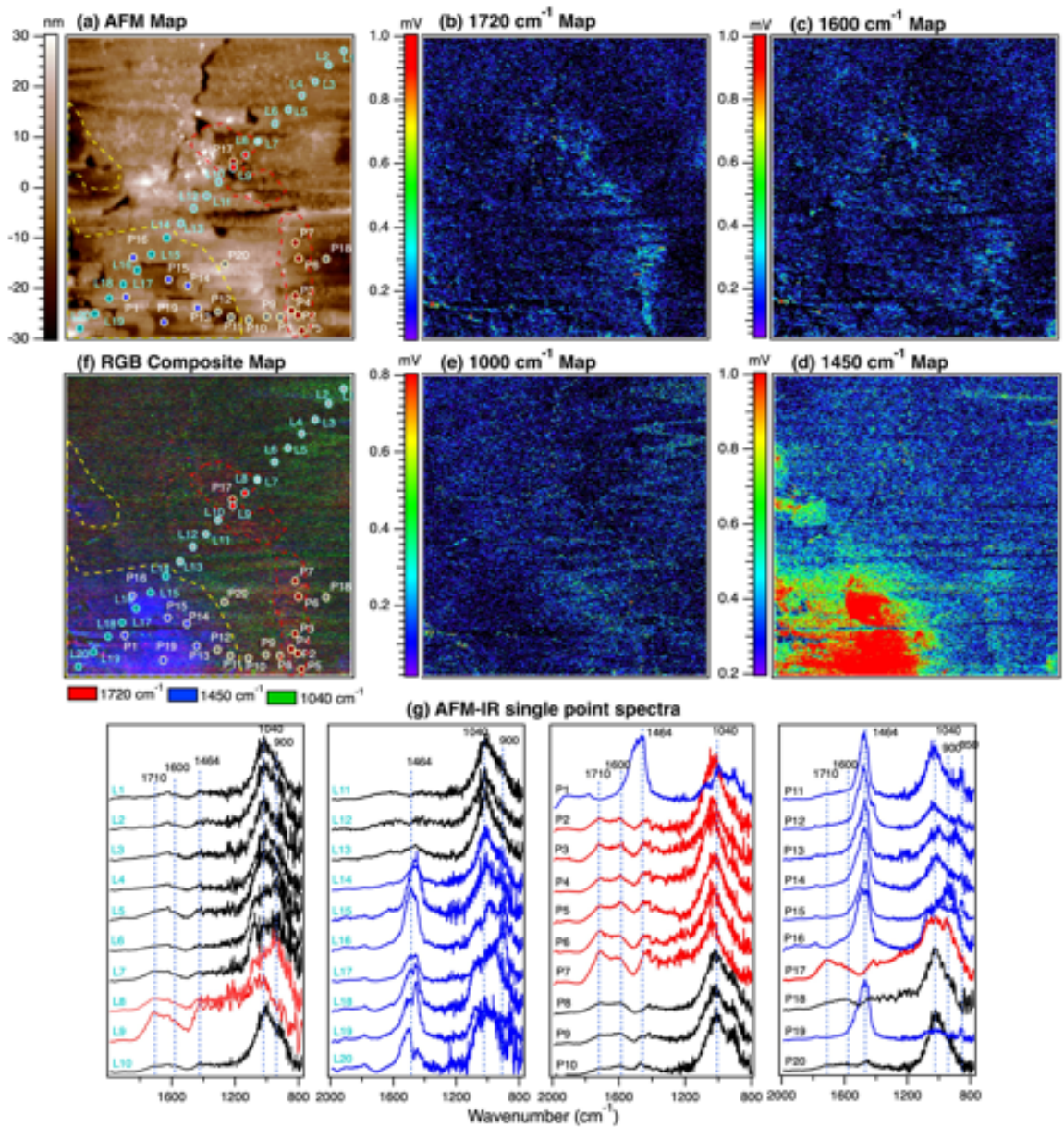

Figure 3. EET 92042 (5 × 5μm$^2$). AFMIR maps recorded from (a) topographical image, at (b) C=O at 1720 cm$^{-1}$, (c) C=C and -OH stretch at 1600 cm$^{-1}$, (d) CO$_3$ at 1450 cm$^{-1}$, (e) silicates Si-O at 1000 cm$^{-1}$ and (f) a composite (red, green, blue) image of the three maps: 1720, 1000 and 1450 cm$^{-1}$, respectively. (g) AFM-IR single point spectra in the 2000 – 700 cm$^{-1}$ range (labelled L1 to L20 with a line of 20 spectra and P1 – P20 with point by point spectra) performed in selected locations shown on the AFM and RGB composite images (a, f).

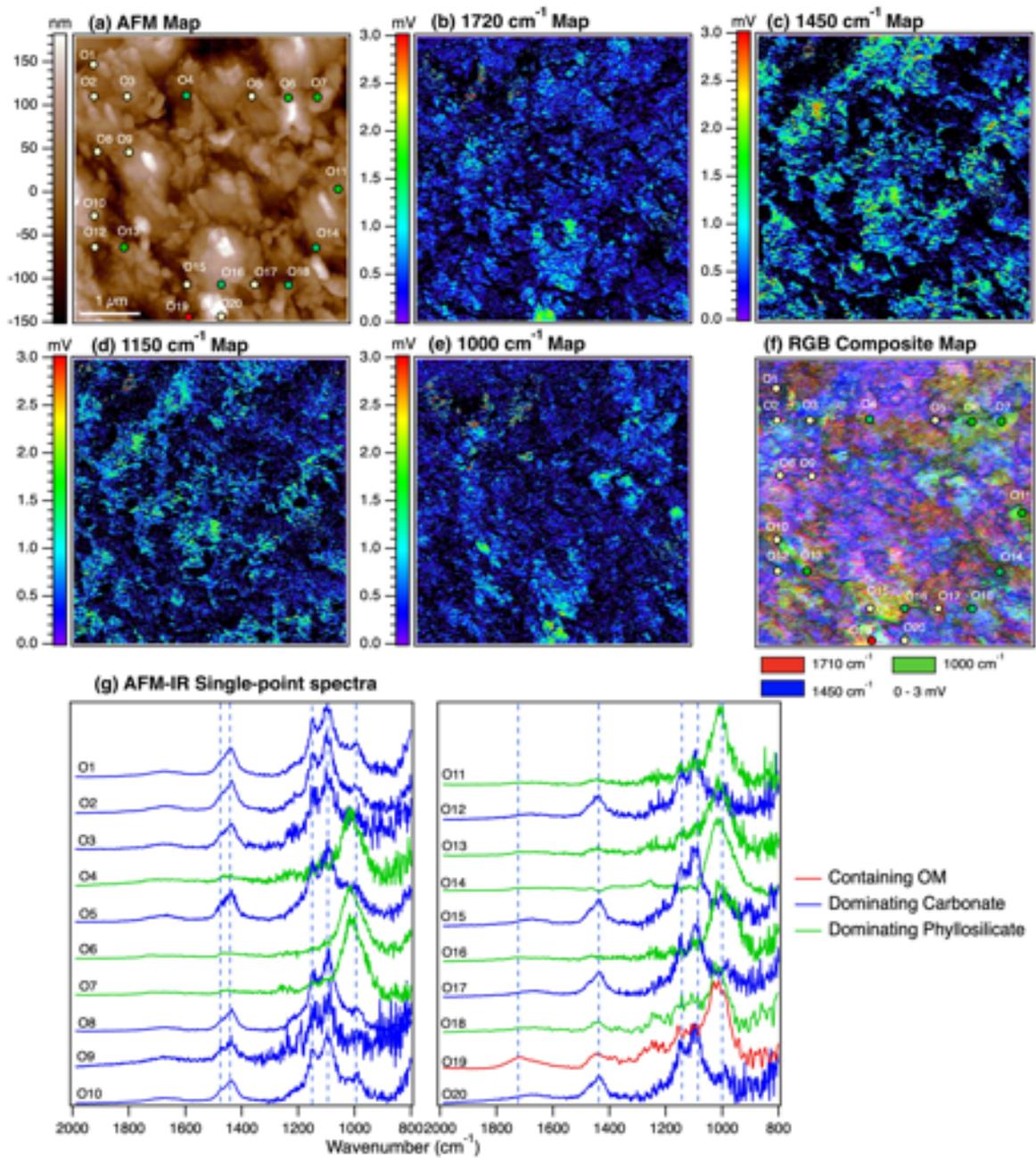

Figure 4. Orgueil – CI Chondrite. (4× 4 μm$^2$). AFM height (a) and AFM-IR images recorded (b) at 1720 cm$^{-1}$, (c) at 1450 cm$^{-1}$, (d) at 1150 cm$^{-1}$, (e) at 1000 cm$^{-1}$ and (f) RGB composite (Red, Green, Blue) image of the three maps: 1720, 1000 and 1450 cm$^{-1}$, respectively. (g) AFM-IR single point spectra in the 2000 – 700 cm$^{-1}$ range (labelled O1 to O20 for point by point spectra) performed in selected locations shown on the AFM and RGB composite images (a, f).

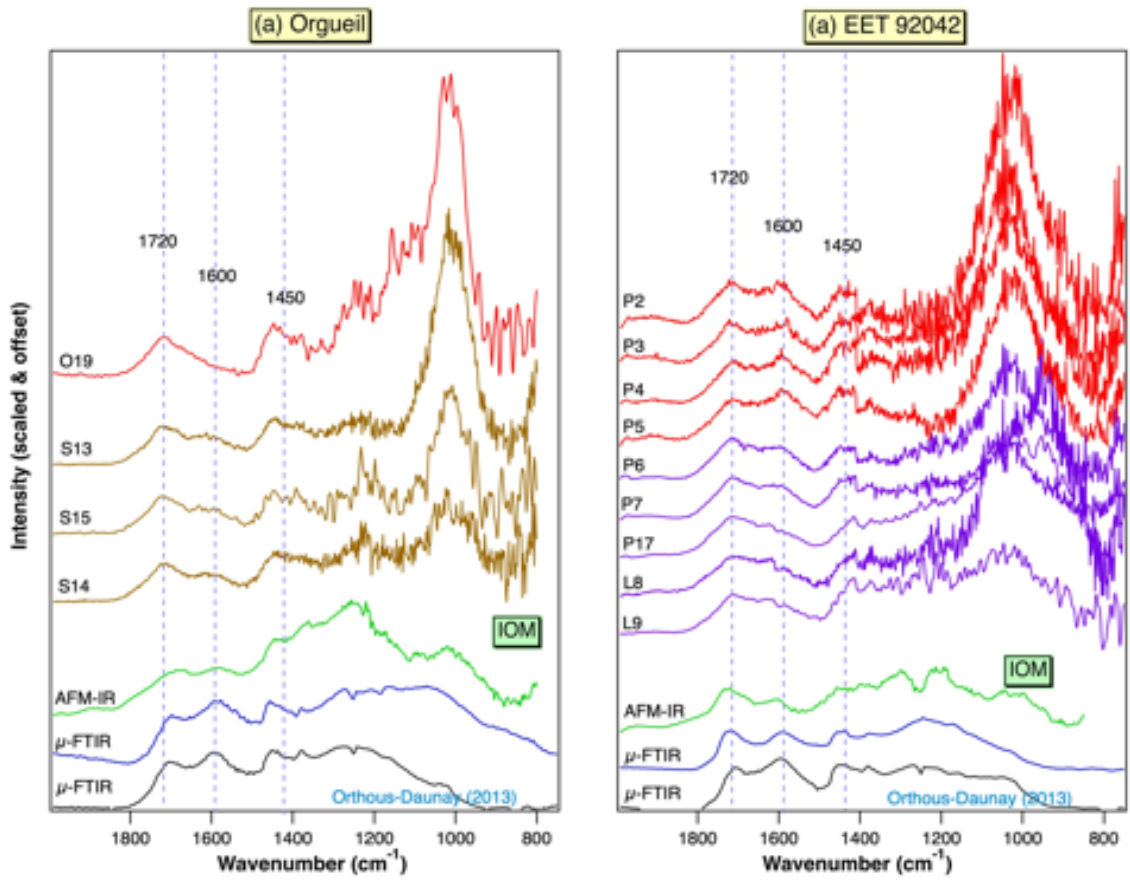

Figure 5. Comparison between infrared spectra of chemically extracted IOM and single-point AFM-IR spectra of organic matter from bulk meteorites (a) Orgueil and (b) EET 92042.

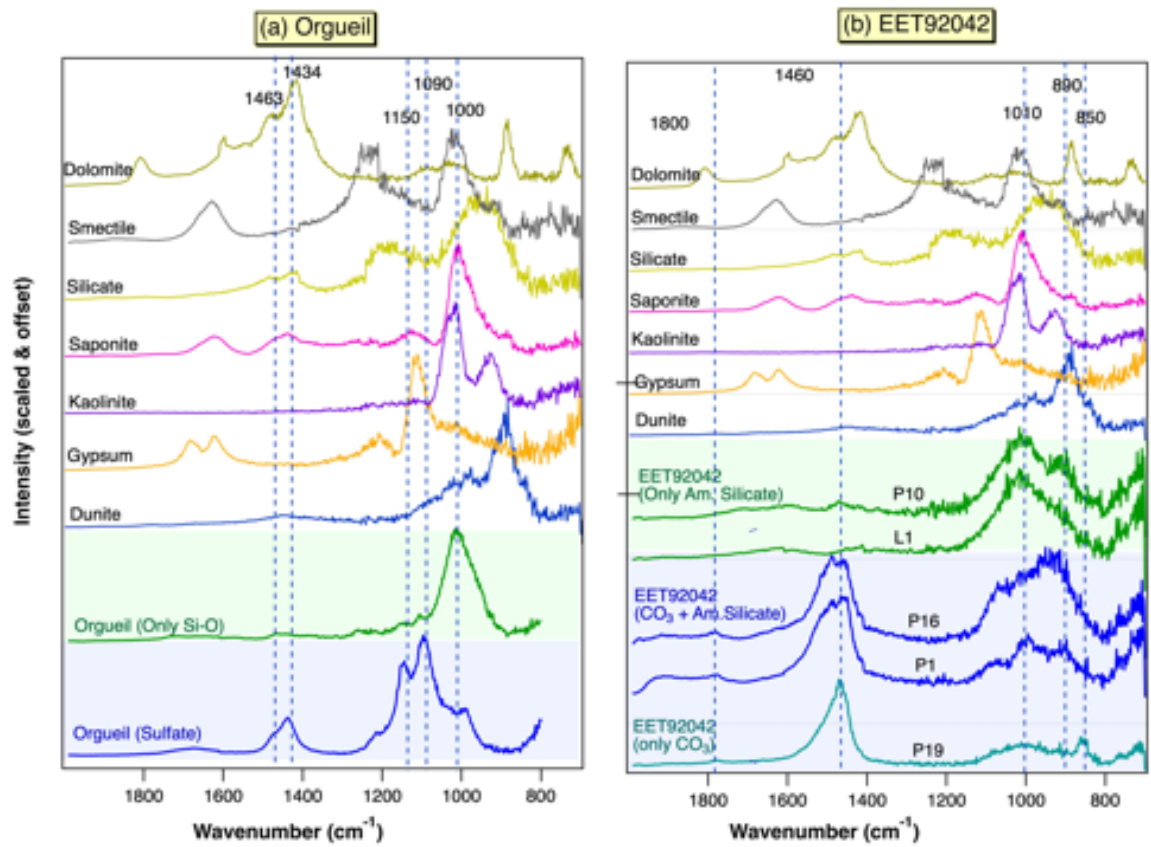

Figure 6. Comparison between pure minerals and single-point AFM-IR spectra from bulk meteorites (a) Orgueil sulfur embedded ultramicrotomy section (b) EET 92042 FIB section

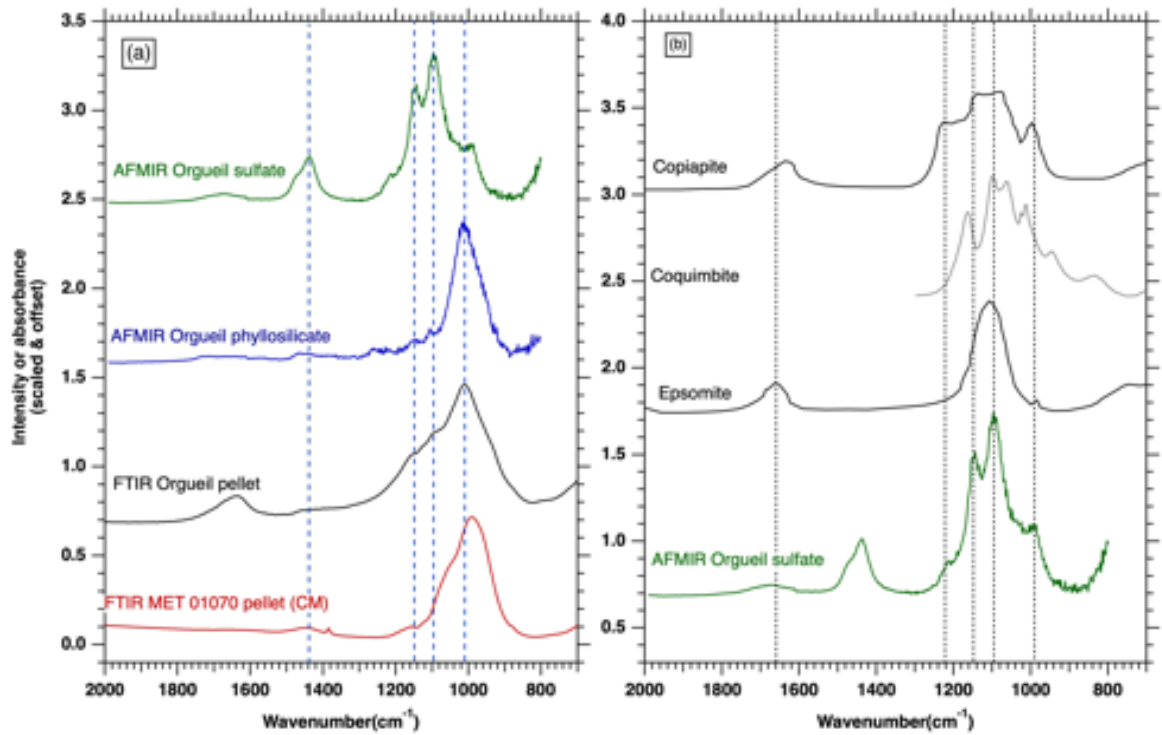

Figure 7. (a) FTIR spectra of the Orgueil CI chondrite and from a typical CM chondrite (Beck et al., 2014) compared to AFM-IR spectra (this study). (b) Comparison of the signature observed in the fine-grained matrix of Orgueil to FTIR spectra of selected sulfate (Fe-rich hydrated: copiapite and coquimbite, and Mg-rich hydrated: epsomite).

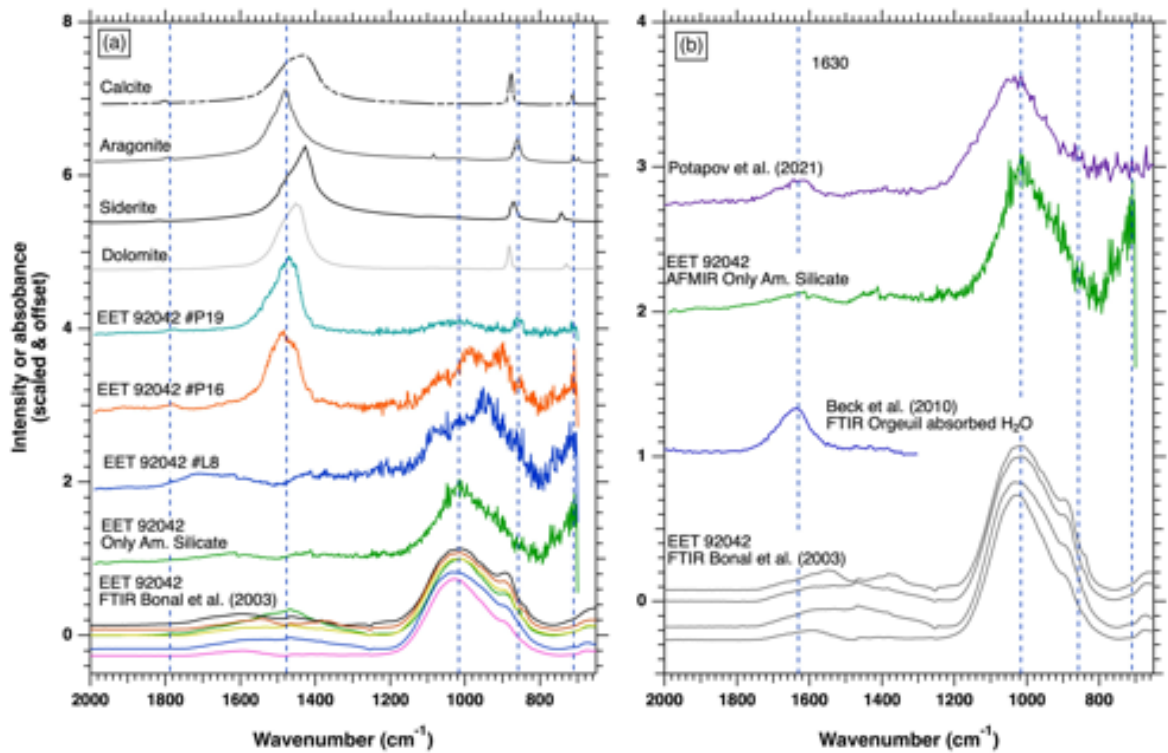

Figure 8. (a) Selected AFM-IR spectra obtained on EET 92042 compared to FTIR measurements from Bonal et al., (2013) on the same meteorite, and transmission spectra of reference carbonates (FTIR) and (b) AFM-IR spectra of hydrated amorphous silicates in EET92042 compared to FTIR spectra of matrix grains (Bonal et al., 2013), Orgueil absorbed water (Beck et al., 2010) and hydrated amorphous silicates from Potapov et al. (2021).